\documentclass[12pt]{article}
\usepackage{amssymb}
\usepackage{amsmath}

\begin{document}

\title{The Density Perturbation Power Spectrum to\\
Second-Order Corrections in the Slow-Roll Expansion}
\author{
Ewan D. Stewart\footnote{ewan@kaist.ac.kr}
\hspace{2cm}
Jin-Ook Gong\footnote{devourer@kaist.ac.kr} \\
{\em Department of Physics, KAIST, Taejeon 305-701, South Korea}}
\date{17th February 2001}
\maketitle

\begin{abstract}
We set up a formalism that can be used to calculate the power spectrum of the curvature perturbations produced during inflation up to arbitrary order in the slow-roll expansion, and explicitly calculate the power spectrum and spectral index up to second-order corrections.

\end{abstract}

\thispagestyle{empty}
\setcounter{page}{0}
\newpage
\setcounter{page}{1}

\section{Introduction}

Curvature perturbations produced during inflation are thought to be the origin of the inhomogeneities necessary for galaxy formation and other large-scale structure.
It will soon be possible to measure the power spectrum of these inhomogeneities very accurately by means of the MAP \cite{map} and {\sc Planck} \cite{planck} satellites and the SDSS program \cite{sdss}.
For example, {\sc Planck} is supposed to be able to measure the power spectrum of the cosmic microwave background radiation with an error of a few percent \cite{planck}, and the spectral index with an error of less than one percent \cite{tegmark}.
Thus, it is very important to calculate the power spectrum precisely in order to get the maximum benefit from these observations.

The power spectrum and spectral index of the curvature perturbations produced during inflation have been calculated up to next-to-leading order, or first-order corrections, in the slow-roll approximation \cite{sr,nakamura}.
Given the current observational constraints on the spectral index, these results for the power spectrum and spectral index are expected to have errors of order a few percent or less and a few tenths of a percent or less, respectively.
These errors are comparable with the expected errors in the planned observations discussed above.
It is thus desirable to improve the accuracy of these calculations to ensure that the calculational errors are small compared with the observational errors.
See Refs.~\cite{further} for further discussion of the accuracy of, and
previous efforts to go beyond, these results.

In this paper, we set up a formalism that can be used to calculate the power spectrum of the curvature perturbations produced during inflation up to arbitrary order in the slow-roll expansion, and explicitly calculate the power spectrum and spectral index up to second-order corrections.
To check our result, we compare it with the two known exact solutions \cite{sr}, power-law inflation and inflation near a maximum.
Our results are only valid for a single component inflaton, but will be generalized to a multi-component inflaton \cite{misao,nakamura} in a forthcoming paper.

\section{The Calculation}

\subsection{Preliminaries}

The scalar linear perturbation to the homogeneous, isotropic background metric is generally expressed as \cite{bardeen}
\begin{equation}
ds^2 = a^2(\eta) \left\{(1+2A)d\eta^2 - 2\partial_iB\,dx^id\eta
- \left[(1+2\mathcal{R})\delta_{ij} + 2\partial_i\partial_jH_T\right]dx^idx^j\right\}.
\end{equation}
It is convenient to express the density perturbation in terms of the intrinsic curvature perturbation of comoving hypersurfaces $\mathcal{R}_\mathrm{c}$ \cite{lukash}, given by
\begin{equation}
\mathcal{R}_\mathrm{c} = \mathcal{R}-\frac{H}{\dot\phi} \delta\phi
\end{equation}
during inflation, where $\delta\phi$ is the perturbation in the inflaton field.
The power spectrum $\mathcal{P}_{\mathcal{R}_\mathrm{c}}(k)$ is given by
\begin{equation}
\langle\mathcal{R}_\mathrm{c}(\mathbf{x},\eta),
\mathcal{R}_\mathrm{c}(\mathbf{y},\eta)\rangle = \int \frac{dk}{k}
\frac{\sin(k|\mathbf{x}-\mathbf{y}|)}{k|\mathbf{x}-\mathbf{y}|}
\mathcal{P}_{\mathcal{R}_\mathrm{c}}.
\end{equation}
The standard result for the power spectrum of the curvature perturbation to leading order in the slow-roll approximation is \cite{ll}
\begin{equation}
\label{sol0}
\mathcal{P}_{\mathcal{R}_\mathrm{c}}(k)
= \left.\left(\frac{H}{2\pi}\right)^2\left(\frac{H}{\dot\phi}\right)^2
\right|_{aH=k}
\end{equation}
and to first-order corrections is \cite{sr}
\begin{equation}
\label{sol1}
\mathcal{P}_{\mathcal{R}_\mathrm{c}}(k)
= \left.\left[1+(4\alpha-2)\epsilon_1+2\alpha\delta_1\right]
\left(\frac{H}{2\pi}\right)^2\left(\frac{H}{\dot\phi}\right)^2\right|_{aH=k},
\end{equation}
where
\begin{equation}\label{eps}
\epsilon_1 = -\frac{\dot H}{H^2}
= \frac{1}{2}\left(\frac{\dot{\phi}}{H}\right)^2, \ \
\delta_1 = \frac{\ddot{\phi}}{H\dot{\phi}} \ ,
\end{equation}
\begin{equation}
\label{alpha}
\alpha \equiv 2-\ln 2 -\gamma \simeq 0.729637
\end{equation}
and $\gamma$ is the Euler-Mascheroni constant, $\gamma\simeq 0.577216$.

Our starting point is the effective action during inflation,
\begin{equation}
S = \int \left[ -\frac{1}{2}R + \frac{1}{2}(\partial_\mu \phi)^2 - V(\phi)
\right] \sqrt{-g} \ d^4 x.
\end{equation}
Defining
\begin{equation}
z \equiv \frac{a\dot \phi}{H}
\ \ \ \mbox{and}\ \ \
\varphi \equiv a\left(\delta\phi-\frac{\dot\phi}{H}\mathcal{R}\right)
= -z\mathcal{R}_\mathrm{c},
\end{equation}
the action for scalar perturbations is \cite{mukhanov}
\begin{equation}
S = \int \frac{1}{2} \left[
\left(\frac{\partial\varphi}{\partial\eta}\right)^2
- \left(\nabla\varphi\right)^2
+ \left(\frac{1}{z}\frac{d^2z}{d\eta^2}\right)\varphi^2
\right] d\eta\,dx^3,
\end{equation}
where the conformal time $\eta$ is given by $d\eta = dt / a $.
Using the Fourier transform of $\varphi$, the equation of motion becomes
\begin{equation}
\label{eqs}
\frac{d^2\varphi_k}{d\eta^2}
+ \left(k^2-\frac{1}{z}\frac{d^2z}{d\eta^2}\right)\varphi_k = 0,
\end{equation}
and $\varphi_k$ satisfies
\begin{equation}\label{bc}
\varphi_k \longrightarrow \left\{
\begin{array}{l l l}
\frac{1}{\sqrt{2k}}e^{-ik\eta} & \mbox{as} & -k\eta \rightarrow \infty \\
A_k z & \mbox{as} & -k\eta \rightarrow 0.
\end{array} \right.
\end{equation}
The first asymptotic solution corresponds to the usual flat space vacuum on scales much smaller than the Hubble distance, and the second is the growing mode on scales much larger than the Hubble distance.
The power spectrum is given by
\begin{equation}\label{ps}
\mathcal{P}_{\mathcal{R}_\mathrm{c}}(k)
= \left(\frac{k^3}{2\pi^2}\right)
\lim_{-k\eta\rightarrow0}\left|\frac{\varphi_k}{z}\right|^2
= \frac{k^3}{2\pi^2}|A_k|^2.
\end{equation}

\subsection{The Formalism of the Calculation}

Our task is to solve Eq.~(\ref{eqs}) with the boundary conditions Eq.~(\ref{bc}) to eventually calculate $A_k$.
We define $y = \sqrt{2k}\, \varphi_k$ and $x = -k\eta$.
Then, the equation of motion for $y$ and the asymptotic solutions are
\begin{equation}
\frac{d^2y}{dx^2} + \left(1-\frac{1}{z}\frac{d^2z}{dx^2}\right)y = 0
\end{equation}
and
\begin{equation}
\label{z-sol}
y \longrightarrow \left\{
\begin{array}{l l l}
e^{ix} & \mbox{as} & x \rightarrow \infty \\
\sqrt{2k}\,A_k z & \mbox{as} & x \rightarrow 0.
\end{array}
\right.
\end{equation}

Now, we can choose the ansatz that $z$ takes the form
\begin{equation}
z = \frac{1}{x}f(\ln x).
\end{equation}
Then we have
\begin{equation}
\frac{1}{z}\frac{d^2z}{dx^2} = \frac{2}{x^2} + \frac{1}{x^2}g(\ln x),
\end{equation}
where
\begin{equation}\label{gf}
g=\frac{-3f'+f''}{f},
\end{equation}
and the equation of motion is
\begin{equation}
\label{exactsol}
\frac{d^2y}{dx^2} + \left(1-\frac{2}{x^2}\right) y = \frac{1}{x^2}g(\ln x) y.
\end{equation}
The homogeneous solution with the correct asymptotic behaviour at $x\rightarrow\infty$ is
\begin{equation}
\label{0sol}
y_0(x) = \left(1 + \frac{i}{x}\right)e^{ix}.
\end{equation}
Therefore, using Green's function, Eq.~(\ref{exactsol}) with the boundary condition Eq.~(\ref{z-sol}) can be written as the integral equation
\begin{equation}
\label{sol}
y(x) = y_0(x) + \frac{i}{2}\int_{x}^{\infty}du \ \frac{1}{u^2} \ g(\ln u) \
y(u) \left[y_0^*(u)y_0(x)-y_0^*(x)y_0(u)\right].
\end{equation}

\subsection{The Slow-Roll Expansion}

We will solve Eq.~(\ref{sol}) perturbatively.
Our small parameters will be the slow-roll \cite{slowroll,sr} parameters
$\epsilon_1$ and $\delta_1$ of Eq.~(\ref{eps}) and
\begin{equation}
\delta_n \equiv \frac{1}{H^n\dot{\phi}}\frac{d^{n+1}\phi}{dt^{n+1}}
\end{equation}
evaluated at horizon crossing ($aH=k$), which we assume to satisfy $\epsilon_1 < \xi$ and $|\delta_n| < \xi^n$ for some small perturbation parameter $\xi$.
Note that the subscript denotes the order in the slow-roll expansion.

We can expand
\begin{equation}\label{zexp}
xz = f(\ln x) = \sum_{n=0}^{\infty} \frac{f_n}{n!}(\ln x)^n,
\end{equation}
where, as we will see, $f_n/f_0$ is of order $n$ in the slow-roll expansion.
This expansion is useful for $\exp(-1/\xi) \ll x \ll \exp(1/\xi)$.

Using
\begin{equation}
x = -k\eta = -k\int\frac{dt}{a} \simeq
\frac{k}{aH}\left(1+\epsilon_1+3\epsilon_1^2+2\epsilon_1\delta_1 \right),
\end{equation}
we can express the $f_n$'s in terms of $\epsilon_1$ and the $\delta_n$ evaluated at $aH=k$.
To leading order in the slow-roll expansion
\begin{equation}
f_2 = \left.\frac{d^2 xz}{(d \ln x)^2}\right|_{x=1}
\simeq \left.\frac{k\dot\phi}{H^2}
\left(8\epsilon_1^2+9\epsilon_1\delta_1+\delta_2\right)
\right|_{aH=k}.
\end{equation}
Similarly, we can obtain
\begin{eqnarray}
f_1 & \simeq & \left.\frac{k\dot\phi}{H^2}\left(-2\epsilon_1-\delta_1-6\epsilon_1^2
-4\epsilon_1\delta_1\right)\right|_{aH=k} \\
f_0 & \simeq & \left. \frac{k\dot\phi}{H^2}\left(1+\epsilon_1+5\epsilon_1^2+3\epsilon_1\delta_1\right)
\right|_{aH=k},
\end{eqnarray}
so we get, up to the appropriate orders for this paper,
\begin{eqnarray}
\label{f_0^{-1}}
\frac{1}{f_0} & \simeq & \left.\frac{H^2}{k\dot\phi}
\left(1-\epsilon_1-4\epsilon_1^2-3\epsilon_1\delta_1\right) \right|_{aH=k} \\
\label{a}
\frac{f_1}{f_0} & \simeq & \left.
-2\epsilon_1-\delta_1-4\epsilon_1^2-3\epsilon_1\delta_1 \right|_{aH=k} \\
\label{b}
\frac{f_2}{f_0} & \simeq & \left.
8\epsilon_1^2+9\epsilon_1\delta_1+\delta_2\right|_{aH=k}.
\end{eqnarray}

Eqs.~(\ref{gf}) and~(\ref{zexp}) give
\begin{equation}\label{gexp}
g(\ln x) = \sum_{n=0}^{\infty} \frac{g_{n+1}}{n!}(\ln x)^n,
\end{equation}
where $g_n$ is of order $n$ in the slow-roll expansion and, up to the appropriate orders,
\begin{eqnarray}\label{g1}
g_1 & \simeq & -3\frac{f_1}{f_0}+\frac{f_2}{f_0} \\
\label{g2}
g_2 & \simeq & 3\left(\frac{f_1}{f_0}\right)^2 - 3\frac{f_2}{f_0}.
\end{eqnarray}

We can expand $y$ as
\begin{equation}\label{yexp}
y(x) = \sum_{n=0}^{\infty}y_n,
\end{equation}
where $y_0(x)$ is the homogeneous solution, Eq.~(\ref{0sol}), and $y_n(x)$ is of order $n$ in the slow-roll expansion.

We can now solve Eq.~(\ref{sol}) perturbatively by substituting Eqs.~(\ref{gexp}) and~(\ref{yexp}) and equating terms of the same order.

\subsection{The Calculation up to Second-Order Corrections}

Substituting $g=g_1$ and $y=y_0$ into the right hand side of Eq.~(\ref{sol}), we can easily find the first order corrections to $y$ as
\begin{eqnarray}\label{fsol}
y_1(x) &=& \frac{i}{2}g_1\int_x^{\infty} du \ \frac{1}{u^2}y_0(u)
\left[y_0^*(u)y_0(x)-y_0^*(x)y_0(u)\right] \nonumber \\
&=& -\frac{1}{3}g_1 \left[ -2ix^{-1}e^{ix} + y_0^*(x)\int_x^{\infty} du \
u^{-1}e^{2iu} \right].
\end{eqnarray}

Now considering up to second-order corrections, substituting $g=g_1+g_2\ln x$ and $y=y_0+y_1$ into Eq.~(\ref{sol}) gives two terms of the same order, due to the product of $g_1$ with $y_1$ and $g_2$ with $y_0$.
Let us call them $y_{21}(x)$ and $y_{22}(x)$, respectively.
Explicitly,
\begin{eqnarray}\label{2sol2}
y_{21}(x) &=& \frac{i}{2}g_1 \int_x^{\infty} du \ \frac{1}{u^2} y_1(u)
\left[y_0^*(u)y_0(x)-y_0^*(x)y_0(u)\right] \nonumber \\
&=& -\frac{i}{9}g_1^2 \left[\frac{2}{3}x^{-1}e^{ix}
+ \left(-\frac{5}{3}x^{-1}+\frac{i}{3}\right)e^{-ix}
\int_x^{\infty} du \ u^{-1}e^{2iu} \right. \nonumber \\
&& \hspace{3em}\left.\mbox{} + iy_0(x) \int_x^{\infty} du \ u^{-1}e^{-2iu}
\int_u^{\infty} dv \ v^{-1}e^{2iv} \right]
\end{eqnarray}
and
\begin{eqnarray}\label{2sol1}
y_{22}(x) &=& \frac{i}{2}g_2 \int_x^{\infty} du \ \frac{1}{u^2} \ln u \
y_0(u)\left[y_0^*(u)y_0(x)-y_0^*(x)y_0(u)\right] \nonumber \\
&=& \frac{i}{3} g_2 \left[x^{-1}\left(\frac{8}{3}+2\ln x\right)e^{ix}
+ \frac{4i}{3} y_0^*(x) \int_x^{\infty}du \ u^{-1}e^{2iu} \right.\nonumber\\
&& \hspace{2em}\left.\mbox{}
+ i y_0^*(x) \int_x^{\infty}du \ u^{-1}\ln u \ e^{2iu} \right].
\end{eqnarray}

Adding together Eqs.~(\ref{0sol}), (\ref{fsol}), (\ref{2sol2}) and~(\ref{2sol1}) gives the perturbative solution of Eq.~(\ref{sol}) for small $\xi|\ln x|$ up to second-order corrections.
It should be accurate for $\exp(1/\xi) \gg x \gg \exp(-1/\xi)$.
Taking the limit $x\rightarrow 0$ with $\xi\ln(1/x)$ fixed and small, we obtain, after some calculation, the asymptotic form for $y$ up to second-order corrections
\begin{eqnarray}\label{final-y}
y(x) & \rightarrow & i \left\{ 1
+ \frac{g_1}{3} \left[\alpha+\frac{i\pi}{2}\right]
+ \frac{g_1^2}{18} \left[\alpha^2-\frac{2}{3}\alpha-4+\frac{\pi^2}{4}
+ i\pi \left(\alpha-\frac{1}{3}\right)\right] \right.\nonumber\\
&& \left.\hspace{1em}\mbox{}
+ \frac{g_2}{6} \left[ \alpha^2+\frac{2}{3}\alpha-\frac{\pi^2}{12}
+ i\pi \left(\alpha+\frac{1}{3}\right) \right] \right\} x^{-1} \nonumber\\
&& \mbox{} - \frac{i}{3} \left\{ g_1
+ \frac{g_1^2}{3} \left[\alpha-\frac{1}{3}+i\frac{\pi}{2}\right]
+ \frac{g_2}{3} \right\} x^{-1}\ln x \nonumber\\
&& \mbox{}
+ \frac{i}{6} \left\{\frac{g_1^2}{3} - g_2\right\} x^{-1}(\ln x)^2.
\end{eqnarray}

The exact asymptotic form for $y$ in the limit $x\rightarrow 0$ is given by Eq.~(\ref{z-sol}).
Expanding this perturbatively as in Eq.~(\ref{zexp}) for small $\xi\ln(1/x)$,
\mbox{i.e.} for x in the range $1 \gg x \gg \exp(-1/\xi)$,
gives the asymptotic form for $y$ up to second-order corrections
\begin{equation}\label{asymp}
y(x) \rightarrow \sqrt{2k}\,A_k f_0 \ x^{-1}
+ \sqrt{2k}\,A_k f_1 \ x^{-1}\ln x
+ \frac{1}{2}\sqrt{2k}\,A_k f_2 \ x^{-1}(\ln x)^2.
\end{equation}
Comparing this with Eq.~(\ref{final-y}),
the coefficient of $x^{-1}$ is the desired result
because it will give $A_k$ up to second-order corrections.
The coefficients of $x^{-1}\ln x$ and $x^{-1}(\ln x)^2$ simply give
the consistent asymptotic behaviour, that is, proportional to $z$.
Therefore, substituting Eqs.~(\ref{g1}) and~(\ref{g2}) into Eq.~(\ref{final-y}), matching the coefficient of $x^{-1}$ with that in Eq.(\ref{asymp}), and using Eq.~(\ref{ps}), the power spectrum is
\begin{equation}\label{amp}
\mathcal{P}_{\mathcal{R}_\mathrm{c}}(k) = \frac{k^2}{(2\pi)^2}\frac{1}{f_0^2}
\left[1 - 2\alpha\frac{f_1}{f_0}
+ \left(3\alpha^2-4+\frac{5\pi^2}{12}\right)\left(\frac{f_1}{f_0}\right)^2
+ \left(-\alpha^2+\frac{\pi^2}{12}\right)\frac{f_2}{f_0}\right].
\end{equation}
Therefore, substituting Eqs.~(\ref{f_0^{-1}}), (\ref{a}) and~(\ref{b})
\begin{eqnarray}\label{final-p}
\mathcal{P}_{\mathcal{R}_\mathrm{c}}(k) & = & \frac{H^4}{(2\pi)^2\dot\phi^2}
\left\{ 1 + \left(4\alpha-2\right)\epsilon_1 + 2\alpha\delta_1
+ \left(4\alpha^2-23+\frac{7\pi^2}{3}\right)\epsilon_1^2
\right.\\ && \left.
+ \left(3\alpha^2+2\alpha-22+\frac{29\pi^2}{12}\right)\epsilon_1\delta_1
+ \left(3\alpha^2-4+\frac{5\pi^2}{12}\right)\delta_1^2
+ \left(-\alpha^2+\frac{\pi^2}{12}\right)\delta_2 \right\} , \nonumber
\end{eqnarray}
where the right hand side should be evaluated at $aH=k$.
The numerical values of the coefficients in this equation are
$0.918549$, $1.459274$, $2.158558$, $4.907929$, $1.709446$, $0.290097$, respectively.
Note the curious cancellations ensuring that the coefficients are of order one.

The spectral index
\begin{equation}
n_{\mathcal{R}_\mathrm{c}}(k)
= 1 + \frac{d \ln P_{\mathcal{R}_\mathrm{c}}}{d \ln k}
\end{equation}
can be easily calculated, and we obtain the result
\begin{eqnarray}\label{n1}
n_{\mathcal{R}_\mathrm{c}}(k)
& = & 1 - 4\epsilon_1 - 2\delta_1
+ (8\alpha-8)\epsilon_1^2 + (10\alpha -6)\epsilon_1\delta_1
- 2\alpha\delta_1^2 + 2\alpha\delta_2 \\
& & + \left(-16\alpha^2+40\alpha-108+\frac{28\pi^2}{3}\right)\epsilon_1^3
+ \left(-31\alpha^2+60\alpha-172+\frac{199\pi^2}{12}\right)\epsilon_1^2\delta_1
\nonumber\\ &&
+ \left(-3\alpha^2+4\alpha-30+\frac{13\pi^2}{4}\right)\epsilon_1\delta_1^2
+ \left(-2\alpha^2 +8-\frac{5\pi^2}{6}\right)\delta_1^3
\nonumber\\ &&
+ \left(-7\alpha^2+8\alpha-22+\frac{31\pi^2}{12}\right)\epsilon_1\delta_2
+ \left(3\alpha^2-8+\frac{3\pi^2}{4}\right)\delta_1\delta_2
+ \left(-\alpha^2 +\frac{\pi^2}{12}\right)\delta_3 \nonumber \,.
\end{eqnarray}
where again the right hand side should be evaluated at $aH=k$.
The numerical values of the coefficients in this equation are
$-2.162903$, $1.296372$, $-1.459274$, $1.459274$,
$4.783868$, $18.945687$, $3.397652$, $-1.289411$,
$5.606983$, $0.999314$, $0.290097$, respectively.

\subsection{Checking against the Exact Solutions}

To check this result, we use the two exact solutions,
power-law inflation \cite{pl} and inflation near a maximum \cite{sr}.
Note that these two solutions can be derived from the Hankel function solution \cite{sr} which corresponds to all orders in $g_1$ with $g_n=0$ for $n>1$.

\subsubsection{Power-law inflation}
In power-law inflation, the potential is
\begin{equation}
V(\phi) = V_0\exp\left(-\sqrt{\frac{2}{p}}\,\phi\right).
\end{equation}
{}From this,
\begin{equation}
\epsilon_1 = \frac{1}{p} \ , \ \ \
\delta_1 = -\frac{1}{p} \ , \ \ \
\delta_2 = 2\delta_1^2 \ , \ \ \
\delta_3 = 6\delta_1^3
\end{equation}
and the power spectrum and the spectral index are \cite{sr}
\begin{eqnarray}
\mathcal{P}_{\mathcal{R}_\mathrm{c}}(k) & = & \frac{pH^2}{8\pi^2}
\left[ 2^{\frac{1}{p-1}} \left(1-\frac{1}{p}\right)^{\frac{p}{p-1}}
\frac{\Gamma(\frac{3}{2}+\frac{1}{p-1})}{\Gamma(\frac{3}{2})} \right]^2
\nonumber\\
& \simeq & \frac{H^4}{(2\pi)^2\dot\phi^2} \left[1+(2\alpha-2)\frac{1}{p}
+\left(2\alpha^2-2\alpha-5+\frac{\pi^2}{2}\right)\frac{1}{p^2}\right]
\end{eqnarray}
and
\begin{equation}
n_{\mathcal{R}_\mathrm{c}} = 1-\frac{2}{p-1}
\simeq 1-\frac{2}{p}-\frac{2}{p^2}-\frac{2}{p^3},
\end{equation}
which agree with Eqs.~(\ref{final-p}) and~(\ref{n1}).

\subsubsection{Inflation near a maximum}
The potential sufficiently near a maximum can be written
\begin{equation}
V(\phi) = V_0\left(1-\frac{1}{2}\mu^2\phi^2\right).
\end{equation}
Therefore, for small $\phi$,
\begin{equation}
\epsilon_1 = 0 \ , \ \ \
\delta_1 = \frac{3}{2}\left(\sqrt{1+\frac{4}{3}\mu^2}-1\right) \ , \ \ \
\delta_2 = \delta_1^2 \ , \ \ \
\delta_3 = \delta_1^3
\end{equation}
and the power spectrum and spectral index are \cite{sr}
\begin{eqnarray}
\mathcal{P}_{\mathcal{R}_\mathrm{c}}(k)
& = & \frac{V_0}{12\pi^2\phi^2\delta_1^2} \left[ 2^{\delta_1}
\frac{\Gamma(\frac{3}{2}+\delta_1)}{\Gamma(\frac{3}{2})} \right]^2
\nonumber\\
& \simeq & \frac{H^4}{(2\pi)^2\dot\phi^2} \left[1+2\alpha\delta_1
+\left(2\alpha^2-4+\frac{\pi^2}{2}\right)\delta_1^2\right],
\end{eqnarray}
and
\begin{equation}
n_{\mathcal{R}_\mathrm{c}} = 1-2\delta_1,
\end{equation}
which again agree with Eqs.~(\ref{final-p}) and~(\ref{n1}).

\subsection{The Spectrum and Spectral Index in terms of the Inflaton Potential}

Now we express every quantity in terms of the potential $V(\phi)$.
Using $3H^2 = V + \frac{1}{2}\dot{\phi}^2$, its derivatives and
$\dot{H} = -\dot{\phi}^2/2$, and defining
\begin{equation}
\mathcal{U}_1 \equiv \left(\frac{V'}{V}\right)^2, \ \
\mathcal{V}_1 \equiv \frac{V''}{V} \ , \ \
\mathcal{V}_2 \equiv \frac{V'V'''}{V^2} \ \ \ \mbox{and} \ \ \
\mathcal{V}_3 \equiv \frac{(V')^2V''''}{V^3} \ ,
\end{equation}
we have
\begin{eqnarray}
H^2 & \simeq & \frac{V}{3} \left[ 1
+ \frac{1}{6}\mathcal{U}_1 - \frac{1}{12}\mathcal{U}_1^2
+ \frac{1}{9}\mathcal{U}_1\mathcal{V}_1 \right]
\\
\left(\frac{H}{\dot\phi}\right)^2 & \simeq & \frac{1}{\mathcal{U}_1}
\left[1 + \frac{2}{3}\mathcal{U}_1 - \frac{2}{3}\mathcal{V}_1
- \frac{4}{9}\mathcal{U}_1^2 + \frac{7}{9}\mathcal{U}_1\mathcal{V}_1
- \frac{1}{9}\mathcal{V}_1^2 - \frac{2}{9}\mathcal{V}_2 \right]
\\
\epsilon_1 & \simeq & \frac{1}{2}\mathcal{U}_1
- \frac{1}{3}\mathcal{U}_1^2 + \frac{1}{3}\mathcal{U}_1\mathcal{V}_1
+ \frac{4}{9}\mathcal{U}_1^3 - \frac{5}{6}\mathcal{U}_1^2\mathcal{V}_1
+ \frac{5}{18}\mathcal{U}_1\mathcal{V}_1^2
+ \frac{1}{9}\mathcal{U}_1\mathcal{V}_2
\\
\delta_1 & \simeq & \frac{1}{2}\mathcal{U}_1 - \mathcal{V}_1
- \frac{2}{3}\mathcal{U}_1^2 + \frac{4}{3}\mathcal{U}_1\mathcal{V}_1
- \frac{1}{3}\mathcal{V}_1^2 - \frac{1}{3}\mathcal{V}_2
\nonumber\\ && \mbox{}
+ \frac{3}{2}\mathcal{U}_1^3 - 4\mathcal{U}_1^2\mathcal{V}_1
+ \frac{23}{9}\mathcal{U}_1\mathcal{V}_1^2 - \frac{2}{9}\mathcal{V}_1^3
+ \frac{17}{18}\mathcal{U}_1\mathcal{V}_2 - \frac{2}{3}\mathcal{V}_1\mathcal{V}_2
- \frac{1}{9}\mathcal{V}_3
\\
\delta_2 & \simeq & \mathcal{U}_1^2 - \frac{5}{2}\mathcal{U}_1\mathcal{V}_1
+ \mathcal{V}_1^2 + \mathcal{V}_2
\nonumber\\ && \mbox{}
- \frac{19}{6}\mathcal{U}_1^3 + \frac{55}{6}\mathcal{U}_1^2\mathcal{V}_1
- \frac{13}{2}\mathcal{U}_1\mathcal{V}_1^2 + \frac{2}{3}\mathcal{V}_1^3
- \frac{5}{2}\mathcal{U}_1\mathcal{V}_2 + 2\mathcal{V}_1\mathcal{V}_2
+ \frac{1}{3}\mathcal{V}_3
\\
\delta_3 & \simeq &
\frac{7}{2}\mathcal{U}_1^3 - \frac{45}{4}\mathcal{U}_1^2\mathcal{V}_1
+ 9\mathcal{U}_1\mathcal{V}_1^2 - \mathcal{V}_1^3
+ 4\mathcal{U}_1\mathcal{V}_2 - 4\mathcal{V}_1\mathcal{V}_2 - \mathcal{V}_3 \,.
\end{eqnarray}
Substituting into Eq.~(\ref{final-p}) gives
\begin{eqnarray}
\mathcal{P}_{\mathcal{R}_\mathrm{c}}(k) & = & \frac{V}{12\pi^2\mathcal{U}_1}
\left[1 + \left(3\alpha-\frac{1}{6}\right)\mathcal{U}_1
+ \left(-2\alpha-\frac{2}{3}\right)\mathcal{V}_1
\right.\nonumber\\ && \left.\mbox{}
+ \left(\frac{3}{2}\alpha^2+\frac{1}{3}\alpha-\frac{77}{6}+\frac{11\pi^2}{8}\right)
\mathcal{U}_1^2
+ \left(-2\alpha^2-\frac{2}{3}\alpha+\frac{142}{9}-\frac{11\pi^2}{6}\right)
\mathcal{U}_1\mathcal{V}_1
\right.\nonumber\\ && \left.\mbox{}
+ \left(2\alpha^2+\frac{2}{3}\alpha-\frac{37}{9}+\frac{\pi^2}{2}\right)
\mathcal{V}_1^2
+ \left(-\alpha^2-\frac{2}{3}\alpha-\frac{2}{9}+\frac{\pi^2}{12}\right)
\mathcal{V}_2 \right].
\end{eqnarray}
The numerical values of the coefficients in this equation are
$2.022245$, $-2.125941$, $1.779141$, $-3.867662$, $2.374857$,
$-0.418550$, respectively.

Using
\begin{equation}
\frac{dX|_{aH=k}}{d\ln k}
\simeq - \left. \left( 1 + \frac{1}{6}\mathcal{U}_1 + \frac{1}{3}\mathcal{V}_1
+ \frac{5}{36}\mathcal{U}_1^2 - \frac{2}{9}\mathcal{U}_1\mathcal{V}_1
+ \frac{2}{9}\mathcal{V}_1^2 + \frac{1}{9}\mathcal{V}_2 \right)
\frac{V'}{V}\frac{dX}{d\phi} \right|_{aH=k}
\end{equation}
or substituting into Eq.~(\ref{n1}) gives
\begin{eqnarray}
n_{\mathcal{R}_\mathrm{c}}(k)
& = & 1 - 3\mathcal{U}_1 + 2\mathcal{V}_1
+ \left(6\alpha-\frac{5}{6}\right)\mathcal{U}_1^2
+ (-8\alpha-1)\mathcal{U}_1\mathcal{V}_1
+ \frac{2}{3}\mathcal{V}_1^2 + \left(2\alpha+\frac{2}{3}\right)\mathcal{V}_2
\nonumber\\ && \mbox{} +
\left(-12\alpha^2+\frac{13}{3}\alpha-\frac{1867}{36}+\frac{11\pi^2}{2}\right)
\mathcal{U}_1^3
+ \left(24\alpha^2+\frac{1}{3}\alpha+\frac{595}{6}-11\pi^2\right)
\mathcal{U}_1^2\mathcal{V}_1
\nonumber\\ && \mbox{}
+ \left(-8\alpha^2-6\alpha-\frac{371}{9}+\frac{14\pi^2}{3}\right)
\mathcal{U}_1\mathcal{V}_1^2
+ \frac{4}{9}\mathcal{V}_1^3
+ \left(-6\alpha^2-2\alpha-\frac{49}{3}+2\pi^2\right)
\mathcal{U}_1\mathcal{V}_2
\nonumber\\ && \mbox{}
+ \left(\alpha^2+\frac{8}{3}\alpha+\frac{28}{3}-\frac{13\pi^2}{12}\right)
\mathcal{V}_1\mathcal{V}_2
+ \left(\alpha^2+\frac{2}{3}\alpha+\frac{2}{9}-\frac{\pi^2}{12}\right)
\mathcal{V}_3 \,.
\end{eqnarray}
The numerical values of the coefficients in this equation are
$3.544490$, $-6.837097$, $2.125941$, $-0.804970$, $3.621120$,
$-3.800854$, $-1.247621$, $1.119331$, $0.418550$, respectively.

\subsection*{Acknowledgements}

We thank Jai-chan Hwang for encouraging us to complete and publish this work,
and Jai-chan Hwang, Kiwoon Choi and Pyungwon Ko for helpful comments.
This work was supported in part by Brain Korea 21 and KRF grant 2000-015-DP0080.

\end{document}